\documentclass[sigconf]{acmart}
\usepackage{booktabs}  % For \toprule, \midrule, \bottomrule
\usepackage{makecell}  % For multi-line cells
\usepackage{amsmath,amsfonts}
\usepackage{algorithmic}
\usepackage{algorithm}
\usepackage{array}
\usepackage[caption=false,font=normalsize,labelfont=sf,textfont=sf]{subfig}
\usepackage{textcomp}
\usepackage{stfloats}
\usepackage{url}
\usepackage{verbatim}
\usepackage{graphicx}
\usepackage{xcolor}
\usepackage{multirow}
\usepackage{lettrine}
\AtBeginDocument{%
  }

\acmYear{2025}\copyrightyear{2025}
\setcopyright{cc}
\setcctype[4.0]{by}
\acmConference[OpenRan '25]{2nd ACM Workshop on Open and AI RAN}{November 4--8, 2025}{Hong Kong, China}
\acmBooktitle{2nd ACM Workshop on Open and AI RAN (OpenRan '25), November 4--8, 2025, Hong Kong, China}
\acmDOI{10.1145/3737900.3770175}
\acmISBN{979-8-4007-1977-6/25/11}

\begin{document}

\title{Poster: Agentic AI meets Neural Architecture Search: Proactive Traffic Prediction for AI-RAN}

\author{Abdelaziz Salama, Mohammed M. H. Qazzaz, Zeinab Nezami, Maryam Hafeez, Syed Ali Raza Zaidi}
% \authornote{Corresponding author: A.M.Salama@Leeds.ac.uk}
\affiliation{%
  \institution{School of Electrical and Electronic Engineering, University of Leeds}
  \city{Leeds}
  \country{United Kingdom}
}


\begin{abstract}
Next-generation wireless networks require intelligent traffic prediction to enable autonomous resource management and handle diverse, dynamic service demands. The Open Radio Access Network (O-RAN) framework provides a promising foundation for embedding machine learning intelligence through its disaggregated architecture and programmable interfaces. This work applies a Neural Architecture Search (NAS)-based framework that dynamically selects and orchestrates efficient Long Short-Term Memory (LSTM) architectures for traffic prediction in O-RAN environments. Our approach leverages the O-RAN paradigm by separating architecture optimisation (via non-RT RIC rApps) from real-time inference (via near-RT RIC xApps), enabling adaptive model deployment based on traffic conditions and resource constraints. Experimental evaluation across six LSTM architectures demonstrates that lightweight models achieve $R^2 \approx 0.91$--$0.93$ with high efficiency for regular traffic, while complex models reach near-perfect accuracy ($R^2 = 0.989$--$0.996$) during critical scenarios. Our NAS-based orchestration achieves a 70-75\% reduction in computational complexity compared to static high-performance models, while maintaining high prediction accuracy when required, thereby enabling scalable deployment in real-world edge environments.
\end{abstract}






\maketitle

\section{Introduction}\label{sec:intro}

The evolution towards next-generation wireless networks demands intelligent and autonomous network management capabilities that extend beyond traditional performance optimisation approaches. Modern telecommunication infrastructures require predictive analytics and proactive resource allocation to handle increasingly complex traffic patterns and diverse service requirements. Within this context, the Open Radio Access Network (O-RAN) framework emerges as a pivotal architecture that enables intelligent network control through disaggregated components and programmable interfaces \cite{polese2023understanding}. The integration of agent AI models, particularly Long Short-Term Memory (LSTM) agents, within O-RAN's RAN Intelligent Controllers (RICs) presents significant opportunities for real-time traffic prediction and autonomous network optimisation, as we have shown in our earlier work \cite{salama2025edge}.

\begin{figure}
    \centering
    \includegraphics[width=1\linewidth]{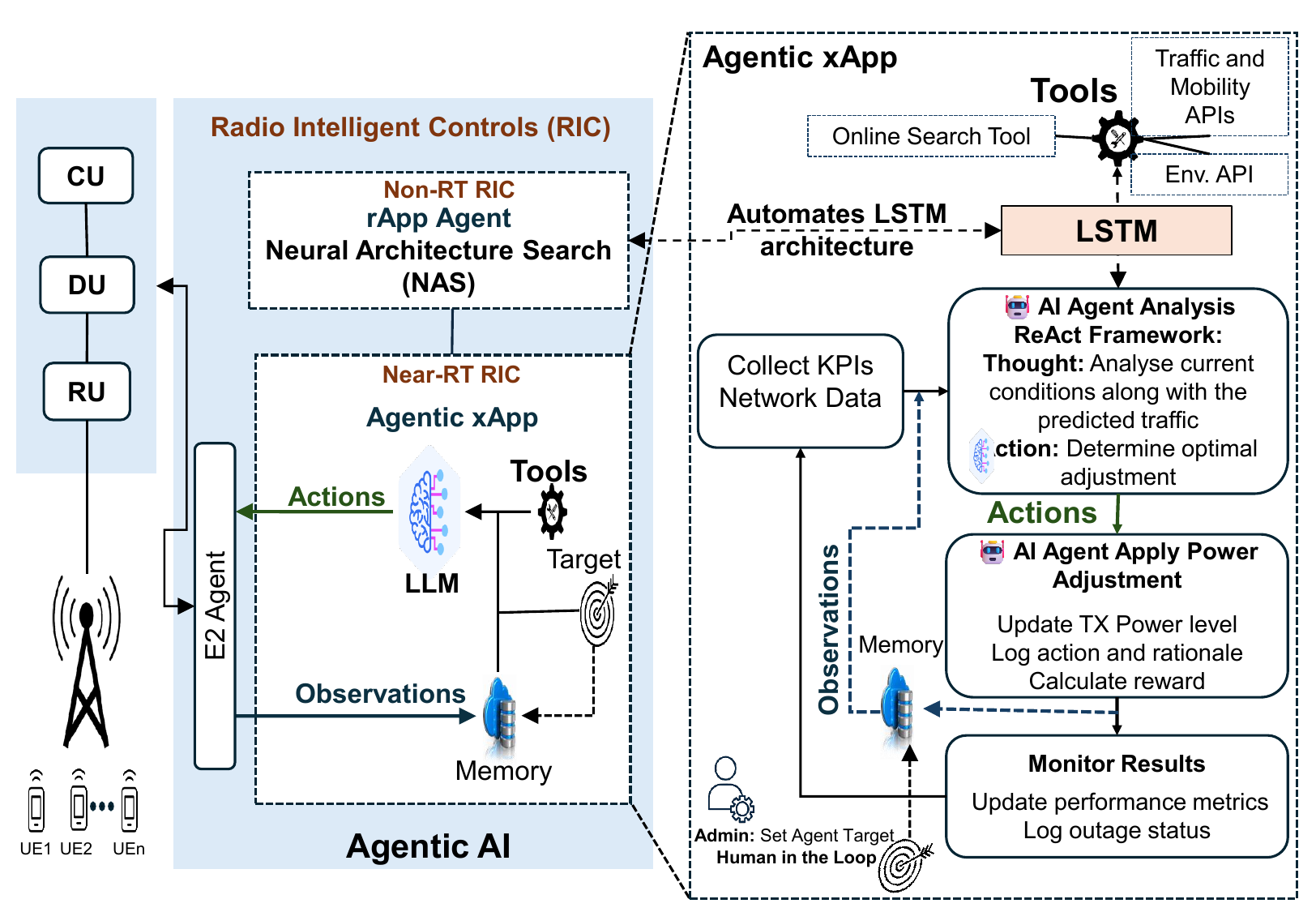}
    \caption{Balanced complexity agentic AI framework using NAS-based LSTM for traffic prediction.}
    \label{fig:NAS-Based-LSTM}
\end{figure}

The deployment of Agentic AI and LSTM-based prediction agents in O-RAN environments faces a fundamental trade-off between predictive accuracy and computational efficiency. Large and complex models deliver high accuracy but are often impractical at the edge due to constraints in computation, memory, and energy, leading to high operational costs and limited scalability \cite{xie2025analysis}. Conversely, overly lightweight models may fail to capture the temporal dynamics of network traffic, resulting in suboptimal prediction performance.

This challenge necessitates an adaptive framework that dynamically selects optimal architectures to balance accuracy and efficiency based on operational context. Leveraging O-RAN's modular architecture, we implement Neural Architecture Search (NAS) \cite{dissem2024neural, elsken2019neural} as an rApp to identify optimal LSTM configurations for time-series traffic prediction, while xApps execute the selected models in real-time. This approach maintains computational efficiency during normal operations while ensuring high accuracy during critical network events.

\section{System Model}\label{sec:methodology}

The proposed framework operates within the Open Radio Access Network (O-RAN) architecture to predict in advance the network traffic load and enable proactive optimisation in telecom networks. The architecture employs a two-tier approach that separates model optimisation from real-time inference execution.

At the architectural optimisation level, a NAS module is deployed as a non-real-time RAN Intelligent Controller (non-RT RIC) rApp, as shown in Fig \ref{fig:NAS-Based-LSTM}. This component periodically explores and evaluates different LSTM configurations to determine the optimal architecture for traffic forecasting based on current network conditions and performance requirements. The rApp operates with relaxed timing constraints, allowing for comprehensive architectural exploration and evaluation.

Once the optimal configuration is identified, it is instantiated as an xApp in the near-real-time RIC, functioning as a low-latency inference agent for real-time traffic prediction. This separation of concerns ensures that the computationally intensive architecture search process does not interfere with the time-critical prediction tasks, while enabling continuous model optimisation as network conditions evolve.

% \subsection{Traffic Prediction Model}

The prediction model leverages historical network data through a timestep input sequence to forecast the next traffic load value using the selected LSTM architecture, ensuring sufficient context while preserving computational efficiency for real-time operation.

The input features combine network KPIs and temporal parameters. Network KPIs capture physical layer conditions that directly influence traffic patterns, while temporal features (hour, day of week, weekend/peak flags) capture cyclical traffic behaviours in telecom networks.

% \subsection{LSTM Architecture and Complexity Analysis}

An LSTM cell regulates information flow through three key gates: the input gate ($i_t$), forget gate ($f_t$), and output gate ($o_t$), which control the cell state and hidden state transitions \cite{hochreiter1997long}. The computational complexity of an LSTM layer with input dimension $d_x$ and hidden dimension $d_h$ is quantified by its parameter count \cite{xie2025analysis}:

\begin{equation}
C_{\text{LSTM}} = 4 \times (d_x d_h + d_h^2 + d_h),
\end{equation}
where the factor $4$ accounts for the input, forget, output, and candidate cell gates. This formulation enables precise complexity assessment for architecture comparison.

The NAS component evaluates six candidate LSTM architectures, systematically exploring the trade-space between predictive accuracy and computational cost. The search space ranges from lightweight single-layer models with 32--64 units, suitable for resource-constrained deployments, to multi-layer configurations with up to 256 units in the first layer, designed for high-accuracy applications.

To enable objective architecture comparison, we define a normalised efficiency metric that balances predictive performance $P$ with computational complexity:

\begin{equation}
E = \frac{P}{C_{\text{norm}}}, \quad 
C_{\text{norm}} = \frac{C_{\text{LSTM}}}{\max(C_{\text{LSTM}})}.
\end{equation}

This formulation allows direct comparison of architectures across varying sizes, explicitly quantifying the trade-off between prediction accuracy and computational cost.

Consequently, the NAS framework evaluates six distinct LSTM architectures designed to explore the trade-off between computational complexity and predictive performance. The \textit{Lightweight} architectures (Lightweight-32 and Lightweight-64) employ single-layer configurations with 32 and 64 hidden units, respectively, utilising a reduced feature set of 6 input parameters to minimise computational overhead for edge deployment scenarios. The \textit{Balanced} architectures (Balanced-Small and Balanced-Medium) implement two-layer LSTM configurations with 64×32 and 100×50 hidden unit arrangements, processing the full 8-feature input set to achieve optimal performance-efficiency trade-offs suitable for production environments. The \textit{Deep-Performance} architecture employs a three-layer configuration (128×100×64 units) with an expanded 10-feature input set, targeting high-accuracy applications where computational resources are less constrained. Finally, the \textit{Ultra-Performance} architecture employs a three-layer design with significantly larger hidden dimensions (512×256×128 units) and a comprehensive 16-feature input set, representing the upper bound of model capacity for scenarios demanding maximum predictive accuracy regardless of computational cost.

\begin{table}[htbp]
  \caption{NAS-based LSTM architecture comparison for O-RAN traffic prediction.}
  \label{tab:arch_comparison}
  \centering
  \resizebox{\columnwidth}{!}{%
    \begin{tabular}{lccccccccc}
      \toprule
      \textbf{Arch} & \textbf{Params} & \textbf{Size} & \textbf{MAE} & \textbf{RMSE} & \textbf{MAPE} &
      \makecell{\textbf{$R^2$} \\ \textbf{(Reg)}} &
      \makecell{\textbf{$R^2$} \\ \textbf{(Crit)}} &
      \makecell{\textbf{$R^2$} \\ \textbf{(Overall)}} &
      \textbf{Eff} \\
      \midrule
      Light-32   & 25K   & 0.02 & 0.006 & 0.018 & 1.95 & 0.976 & 0.860 & 0.934 & 0.597 \\
      Light-64   & 38K   & 0.07 & 0.004 & 0.015 & 1.63 & 0.980 & 0.895 & 0.914 & 0.496 \\
      Bal-Small  & 44K   & 0.17 & 0.007 & 0.016 & 2.53 & 0.981 & 0.910 & 0.949 & 0.903 \\
      Bal-Med    & 74K   & 0.28 & 0.008 & 0.017 & 4.29 & 0.986 & 0.965 & 0.975 & 0.950 \\
      Deep-Perf  & 205K  & 0.78 & 0.009 & 0.019 & 4.91 & 0.990 & 0.970 & 0.989 & 0.901 \\
      Ultra-Perf & 1.08M & 4.13 & 0.008 & 0.019 & 4.03 & 0.996 & 0.982 & 0.996 & 0.279 \\
      \bottomrule
    \end{tabular}%
  }
\end{table}

\begin{figure*}[htbp]
\centering
\includegraphics[width=0.69\textwidth, height=0.323\textheight]{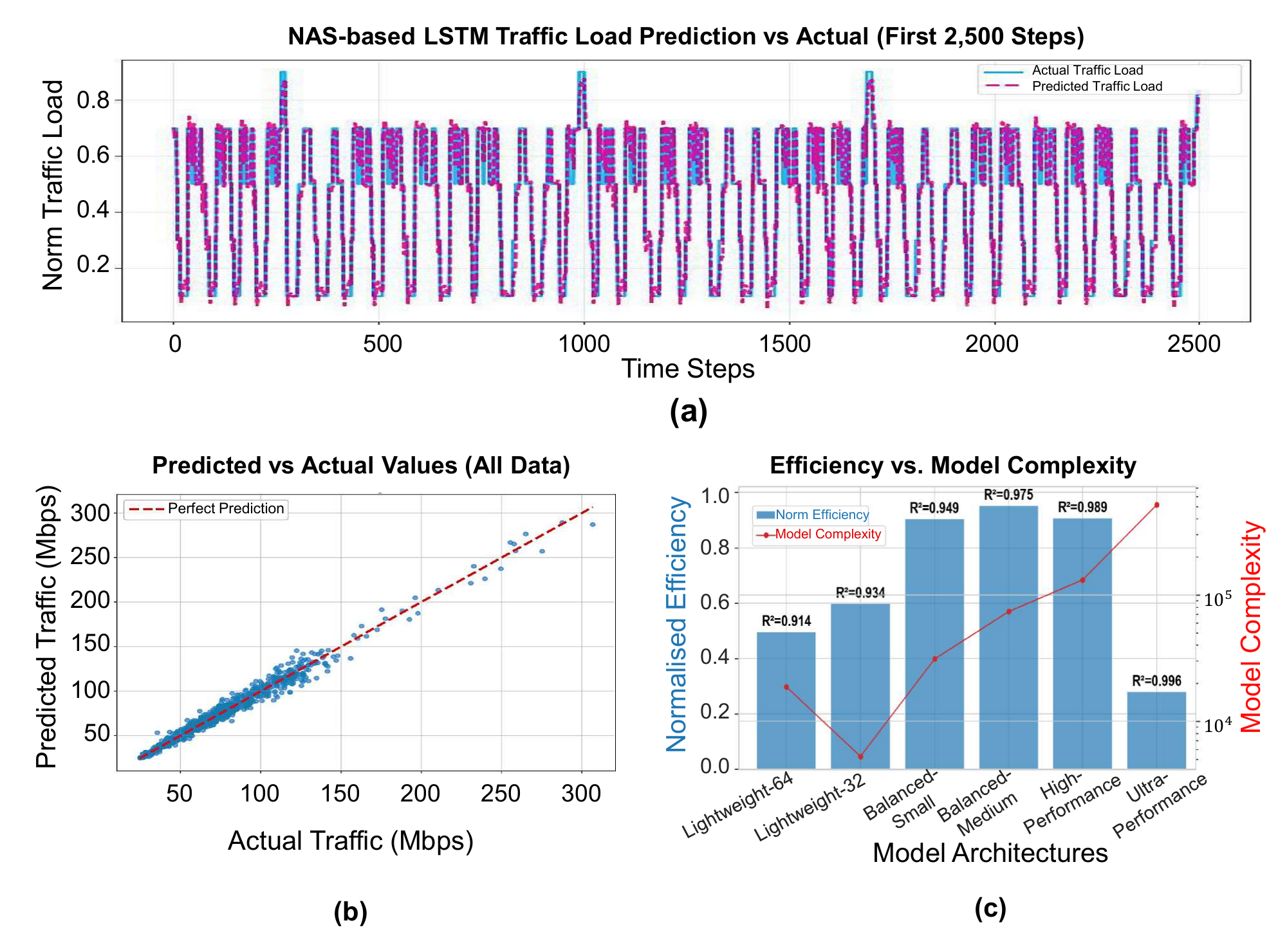}
\caption{NAS-based LSTM traffic prediction results: (a) time series prediction over 2,500 steps, (b) predicted vs actual traffic correlation, and (c) architecture efficiency analysis across model complexities.}
\label{fig:load_balance_improvement}
\end{figure*}

\section{Results and Discussion}\label{sec:results}

The system employs a multi-dimensional evaluation approach that considers both predictive accuracy and computational efficiency. Accuracy is assessed using standard regression metrics, including Mean Absolute Error (MAE), Root Mean Squared Error (RMSE), Coefficient of Determination ($R^2$), and Mean Absolute Percentage Error (MAPE). Efficiency is evaluated in terms of model size, parameter count, and the performance-to-complexity ratio, highlighting architectures that satisfy network requirements while minimising computational overhead.

Within the proposed NAS-based LSTM framework, each LSTM model functions as an xApp agent in the near-real-time RIC, tasked with mobile network traffic prediction. These models process data from two primary sources: (i) historical time series of traffic load and network parameters, and (ii) dynamic contextual information obtained through external APIs. API integration ensures context-awareness by incorporating real-time updates, such as user mobility, service demand, or unexpected network events. Consequently, predictions are not solely based on past patterns but adapt to evolving operational conditions, enhancing robustness under critical scenarios.

The evaluation of our NAS-based LSTM architecture reveals a trade-off between model complexity, computational efficiency, and predictive performance (Table~\ref{tab:arch_comparison}). Lightweight models (e.g., Lightweight-3 and Lightweight-6) achieve low error values (MAE $\approx 0.004$--$0.006$) with overall $R^2$ scores around $0.91$--$0.93$, while maintaining high efficiency in regular traffic scenarios and minimal parameter counts. Balanced architectures, such as Balanced-Small and Balanced-Medium, improve prediction accuracy ($R^2 = 0.949$ and $0.975$, respectively) with moderate complexity increases. Larger models, including Deep-Performance and Ultra-Performance, attain near-perfect $R^2$ values ($0.989$ and $0.996$), particularly under critical traffic conditions, outperforming smaller models ($R^2 = 0.97$ and $0.982$). These gains, however, come at the cost of higher computational demand and reduced efficiency. Notably, under regular traffic, all models perform comparably well ($R^2 > 0.97$), indicating that the benefit of complex models is most pronounced during critical scenarios.

The proposed NAS-based agent dynamically orchestrates multiple LSTM architectures, from lightweight to ultra-performance designs, each instantiated as an xApp variant tailored to operator requirements. Lightweight models are optimised for efficiency and minimal resource use, making them suitable for regular traffic or resource-constrained edge environments. Intermediate architectures leverage broader features from APIs and historical data to enhance prediction accuracy with moderate cost. Larger architectures exploit extended temporal dependencies and richer context to maximise accuracy during sudden traffic surges or anomalies.

In practice, our framework predominantly adopts the Balanced-Medium model for regular scenarios. This choice keeps the system proactive while reducing computational complexity by approximately 70--75\% compared to Deep- and Ultra-Performance models, which are deployed selectively for high-impact events. This hierarchical and adaptive orchestration enables the O-RAN framework to flexibly assign the most appropriate xApp based on network conditions and resource availability, ensuring lightweight and intermediate models perform reliably under normal conditions, while larger models handle extreme scenarios efficiently.

\section{Conclusion}\label{sec:conclusion}
This work proposes a NAS-based LSTM framework for O-RAN traffic prediction that balances predictive accuracy and computational efficiency in edge environments. It dynamically orchestrates LSTM models ranging from lightweight (25K params, $R^2 \approx 0.91$--$0.93$) to ultra-performance (1.08M params, $R^2 = 0.996$). A key innovation is separating architecture search (rApps) from real-time inference (xApps), allowing adaptive model selection based on network context and resource availability.

The system achieves optimal efficiency by primarily deploying Balanced-Medium models ($R^2 = 0.975$) during normal traffic, while activating heavier models for critical events. This reduces computational load by 70--75\% compared to static high-performance approaches, with no significant loss in accuracy. Additionally, the integration of API-driven context awareness further enhances model robustness across diverse operational conditions.

\section*{Acknowledgment}
This research was funded by EPSRC CHEDDAR (EP/X040518/1, EP/Y037421/1), UKRI Grant EP/X039161/1, and MSCA Horizon EU Grant 101086218. Additionally, this research was supported by the UKRI Funding Service under Award UKRI851: Strategic Decision-Making and Cooperation among AI Agents: Exploring Safety and Governance in Telecom.

% \bibliographystyle{ACM-Reference-Format}
\bibliographystyle{unsrt}
\bibliography{Reference}


\end{document}